\documentclass[11pt,a4paper]{article}

\usepackage{subfigure,jcappub,amsfonts,amsmath,amssymb,bm,graphicx}

\usepackage[active]{srcltx}

\title{Neutrino scattering off a black hole surrounded by a magnetized accretion disk}

\author{Maxim Dvornikov}
\emailAdd{maxdvo@izmiran.ru}
\affiliation{Pushkov Institute of Terrestrial Magnetism, Ionosphere and Radiowave Propagation \\ (IZMIRAN),
108840 Moscow, Troitsk, Russia}

\abstract{
We study the neutrino scattering off a rotating black hole with a realistic accretion disk permeated by an intrinsic magnetic field. Neutrino trajectories in curved spacetime as well as the particle spin evolution in dense matter of an accretion disk and in the magnetic field are accounted for exactly. We obtain the fluxes of outgoing ultrarelativistic neutrinos taking into account the change of the neutrino polarization owing to spin oscillations. Using the conservative value of the neutrino magnetic moment and realistic radial distributions of the matter density and the magnetic field strength, we get that these fluxes are reduced by several percent compared to the case when no spin oscillations are accounted for. In some situations, there are spikes in the neutrino fluxes because of the neutrino interaction with the rotating plasma of an accretion disk. Taking into account the uncertainties in the astrophysical neutrino fluxes, the predicted effects turn out to be quite small to be observed with the current neutrino telescopes.
}

\arxivnumber{2102.00806}

\begin{document}

\maketitle

\section{Introduction}

After numerous successful neutrino oscillations experiments (see,
e.g., refs.~\cite{Ace19,Ker20,Aga19}), we believe that neutrinos
are massive particles and there is a mixing between different neutrino
types. These neutrino characteristics are the indications to physics
beyond the standard model. It is also known~\cite{BalKay18} that
neutrino interaction with external fields can modify the process of
neutrino oscillations. For example, the neutrino interaction with
background matter results in the amplification of the transition probability
of neutrino flavor oscillations, which, in its turn, is the most plausible
solution to the solar neutrino problem~\cite{Chr21}. Neutrino interaction
with an electromagnetic field results in transitions between neutrino
states belonging to different helicities, i.e. in spin and/or spin-flavor
oscillations~\cite{Giu19}.

In the present work, we study neutrino spin oscillations, i.e. the
transitions $\nu_{\mathrm{L}}\to\nu_{\mathrm{R}}$ within the same
neutrino flavor under the influence of external fields. Neutrinos
are produced as left polarized particles in the standard model. If
the neutrino helicity changes while a particle propagates, a detector
registers the reduced neutrino flux. As a rule, neutrino spin oscillations
happen in a magnetic field owing to the presence of a nonzero magnetic
moment~\cite{Giu19}. Thus, we suppose that neutrinos are Dirac particles.
Despite the significant experimental progress (see, e.g., ref.~\cite{Alv19}),
the question on the neutrino nature is still open~\cite{Bil20}.

The gravitational interaction also can induce the change of the polarization
of a spinning particle~\cite{Pap51}, including that of neutrinos~\cite{PirRoyWud96}.
The evolution of a fermion spin in curved spacetime under the various
external fields was examined in multiple works (see, e.g., refs.~\cite{SorZil07,ObuSilTer17}).
We studied neutrino spin oscillations in curved spacetime in refs.~\cite{Dvo06,Dvo19}
applying different approaches. In refs.~\cite{Dvo06,Dvo19}, we considered
the neutrino spin evolution both in static gravitational backgrounds,
like that of a black hole (BH), and in time dependent gravitational
fields, such as gravity waves.

In the present work, we discuss spin effects in the neutrino scattering
off BH. This problem is important since we can fix neutrino helicities
for in- and out-states because they are in the flat
spacetime asymptotically. Thus, spin oscillations in the neutrino gravitational scattering
is a more realistic problem compared to the case when a neutrino is
captured gravitationally by BH~\cite{SorZil07,Dvo06}. Note that
the spin-flip of neutrinos in their gravitational scattering was studied
in ref.~\cite{Sor12} using the weak field approximation.

The study of the present work is motivated by the recent observation
of the event horizon silhouette of a supermassive BH (SMBH)~\cite{Aki19}.
An accretion disk, which typically surrounds BH, is a source of photons,
which form its visible image. The same disk can emit neutrinos (see,
e.g., ref.~\cite{KawMin07}), which are gravitationally lensed and
potentially observed by a neutrino telescope. One expects not only
the gravitational lensing of these neutrinos, but also their spin
precession in strong external fields in the vicinity of BH. The
latter effect decreases the observed neutrino flux. It should be noted that the neutrino emission by an accretion disk is effective if the matter density is high~\cite{KawMin07}: $\rho \sim (10^{11} - 10^{12})\,\text{g}\cdot\text{cm}^{-3}$. This scenario is implemented neither for SMBH in our galaxy (Sgr A*) nor in the center of M87. It can happen for BH with a stellar mass entering in a binary system.

Moreover, we can
imagine that a supernova explodes in our galaxy. The flux of emitted
neutrinos is lensed gravitationally by SMBH in the center of the Galaxy
before particles arrive to the Earth. Spin oscillations
of these neutrinos will modify the neutrino flux observed in a neutrino
telescope. Although this situation is very improbable, the possibility of lensing of such neutrinos is discussed in ref.~\cite{MenMocQui06}.

This work is organized in the following way. In section~\ref{sec:SPINEVOL},
we derive the main equations which the neutrino spin obeys while a
particle interacts with external fields in curved spacetime. We adapt
them for the problem of the neutrino scattering off BH. The effective
Schr\"odinger equation for neutrino spin oscillations is derived. Then,
in section~\ref{sec:APPL}, we fix the parameters of a neutrino and
SMBH, which correspond to a realistic situation. Finally, in section~\ref{sec:RES},
we present the neutrino fluxes, which are measured by a terrestrial
detector, obtained in the numerical solution of the Schr\"odinger equation
derived in section~\ref{sec:SPINEVOL}. These fluxes account for the
neutrino interaction with an accretion disk and the magnetic field
in curved spacetime. The derivation of the electromagnetic field in
the vicinity of BH is briefly outlined in appendix~\ref{sec:EMBH}.

\section{Spin evolution in the neutrino scattering\label{sec:SPINEVOL}}

In this section, we derive the effective Hamiltonian for the spin
evolution of a neutrino scattered off a rotating BH surrounded by
an accretion disk and embedded to the external electromagnetic field
$F_{\mu\nu}$, which is defined in world coordinates. The electromagnetic
interaction is owing to the presence of a nonzero neutrino magnetic
moment. The neutrino electroweak interaction with background matter
is in the forward scattering approximation. We consider only the equatorial
neutrino motion.

The metric of spacetime of a rotating BH has the form~\cite{Rez16},
\begin{equation}\label{eq:Kerrmetr}
  \mathrm{d}s^{2}=g_{\mu\nu}\mathrm{d}x^{\mu}\mathrm{d}x^{\nu}=
  \left(
    1-\frac{rr_{g}}{\Sigma}
  \right)
  \mathrm{d}t^{2}+2\frac{rr_{g}a\sin^{2}\theta}{\Sigma}\mathrm{d}t\mathrm{d}\phi-\frac{\Sigma}{\Delta}\mathrm{d}r^{2}-
  \Sigma\mathrm{d}\theta^{2}-\frac{\Xi}{\Sigma}\sin^{2}\theta\mathrm{d}\phi^{2},
\end{equation}
where
\begin{equation}\label{eq:ingmet}
  \Delta=r^{2}-rr_{g}+a^{2},
  \quad
  \Sigma=r^{2}+a^{2}\cos^{2}\theta,
  \quad
  \Xi=
  \left(
    r^{2}+a^{2}
  \right)
  \Sigma+rr_{g}a^{2}\sin^{2}\theta.
\end{equation}
We use the Boyer-Lindquist coordinates $x^{\mu}=(t,r,\theta,\phi)$
in eqs.~(\ref{eq:Kerrmetr}) and~(\ref{eq:ingmet}). We suppose
that the Newton constant equals unity. In this units, the mass of
BH is $M=r_{g}/2$ and its angular momentum $J=Ma$. Here $r_{g}$
is the Schwarzschild radius. The parameter $a\leq r_{g}/2$.

According to refs.~\cite{Dvo06,PomKhr98}, the evolution of a neutrino
spin is defined in the locally Minkowskian frame $x^{a}=e_{\ \mu}^{a}x^{\mu}$,
where $e_{\ \mu}^{a}=\partial x^{a}/\partial x^{\mu}$, $a=0,\dots,3$,
are the vierbein vectors. These vectors are chosen in such a way to
diagonalize the metric in eq.~(\ref{eq:Kerrmetr}), $g_{\mu\nu}=e_{\ \mu}^{a}e_{\ \nu}^{b}\eta_{ab}$,
where $\eta_{ab}=\text{diag}\left(1,-1,-1,-1\right)$ is the Minkowski
metric tensor. One can check by the direct calculation that the following
vectors:
\begin{align}
  e_{0}^{\ \mu}= & \left(\sqrt{\frac{\Xi}{\Sigma\Delta}},0,0,\frac{arr_{g}}{\sqrt{\Delta\Sigma\Xi}}\right),
  \quad
  e_{1}^{\ \mu}= \left(0,\sqrt{\frac{\Delta}{\Sigma}},0,0\right),\nonumber \\
  e_{2}^{\ \mu}= & \left(0,0,\frac{1}{\sqrt{\Sigma}},0\right),
  \quad
  e_{3}^{\ \mu}= \left(0,0,0,\frac{1}{\sin\theta}\sqrt{\frac{\Sigma}{\Xi}}\right),\label{eq:vierbKerr}
\end{align}
satisfy the relation $\eta_{ab}=e_{a}^{\ \mu}e_{b}^{\ \nu}g_{\mu\nu}$.

The four vector of the spin $s^{a}=e_{\ \mu}^{a}S^{\mu}$, where $S^{\mu}$
is the spin vector in world coordinates, in the locally Minkowskian
frame for a neutrino interacting with the gravitational and electromagnetic
fields, as well as with a background matter, obeys the equation~\cite{Dvo13},
\begin{equation}\label{eq:BMTvierb}
  \frac{\mathrm{d}s^{a}}{\mathrm{d}t}=\frac{1}{U^{t}}
  \left[
    G^{ab}s_{b}+2\mu
    \left(
      f^{ab}s_{b}-u^{a}u_{b}f^{bc}s_{c}
    \right)+
    \sqrt{2}G_{\mathrm{F}}\varepsilon^{abcd}g_{b}u_{c}s_{d}
  \right],
\end{equation}
where $G_{ab}=\gamma_{abc}u^{c}$ is the antisymmetric tensor accounting
for the gravitational interaction of neutrinos, $\gamma_{abc}=\eta_{ad}e_{\ \mu;\nu}^{d}e_{b}^{\ \mu}e_{c}^{\ \nu}$
are the Ricci rotation coefficients, the semicolon stays for the covariant
derivative, $g^{a}=e_{\ \mu}^{a}G^{\mu}=(g{}^{0},\mathbf{g})$ is
the effective potential of the neutrino matter interaction in the
locally Minkowskian frame, $f_{ab}=e_{a}^{\ \mu}e_{b}^{\ \nu}F_{\mu\nu}=(\mathbf{e},\mathbf{b})$
is the electromagnetic field tensor in the locally Minkowskian frame,
$u^{a}=(u^{0},\mathbf{u})=e_{\ \mu}^{a}U^{\mu}$, $U^{\mu}= \left( U^{t},U^{r},U^{\theta},U^{\phi} \right) = \mathrm{d}x^{\mu}/\mathrm{d}s$
is the neutrino velocity in the world coordinates, $\varepsilon^{abcd}$
is the absolute antisymmetric tensor in Minkowski spacetime having
$\varepsilon^{0123}=+1$, $\mu$ is the neutrino magnetic moment,
and $G_{\mathrm{F}}=1.17\times10^{-5}\,\text{GeV}^{-2}$ is the Fermi
constant.

We suppose that the background matter is an electroneutral hydrogen
plasma with $n_{e}=n_{p}$, where $n_{e,p}$ are the number densities
of electrons and protons. It is assumed to be unpolarized and move
as a whole. In this case~\cite{DvoStu02}, $G^{\mu}=n_{e}U_{f}^{\mu}$,
where $U_{f}^{\mu}$ is the macroscopic four velocity of plasma defined
in the world coordinates. Here $n_{e}$ is the invariant electron
number density given in the rest frame of plasma.

We consider plasma forming an accretion disk, in which particles move
around BH on circular orbits with the radius $r$. In this case, $U_{f}^{r}=0$
and $U_{f}^{\theta}=0$. Using the results of ref.~\cite{BarPreTeu72},
one gets for the remaining nonzero components,
\begin{equation}\label{eq:Uftr}
  U_{f}^{t}=\frac{\sqrt{2}x^{3/2}+\lambda z}{\sqrt{2x^{3}-3x^{2}+2\lambda\sqrt{2}zx^{3/2}}},
  \quad
  U_{f}^{\phi}=\frac{1}{r_{g}}\frac{\lambda}{\sqrt{2x^{3}-3x^{2}+2\lambda\sqrt{2}zx^{3/2}}},
\end{equation}
where $x=r/r_{g}$ and $z=a/r_{g}$. The parameter $\lambda$ corresponds
to a disk, which corotates ($\lambda=+1$) or counter-rotates ($\lambda=-1$)
BH.

Instead of the four vector $s^{a}$, we deal with the invariant three
vector $\bm{\bm{\zeta}}$ of the neutrino spin, which describes the
polarization in the neutrino rest frame. These spin vectors are related
by the following expression:
\begin{equation}\label{eq:nuspinzeta}
  s^{a}=
  \left(
    (\bm{\bm{\zeta}}\cdot\mathbf{u}),\bm{\zeta}+\frac{\mathbf{u}(\bm{\bm{\zeta}}\cdot\mathbf{u})}{1+u^{0}}
  \right).
\end{equation}
Using eqs.~(\ref{eq:BMTvierb}) and~(\ref{eq:Uftr}), we get the
equation for $\bm{\bm{\zeta}}$ in the form,
\begin{equation}\label{eq:nuspinrot}
  \frac{\mathrm{d}\bm{\bm{\zeta}}}{\mathrm{d}t}=2(\bm{\bm{\zeta}}\times\bm{\bm{\Omega}}),
\end{equation}
where
\begin{align}\label{eq:vectG}
  \bm{\bm{\Omega}} = &
  \frac{1}{U^{t}}
  \bigg\{
    \frac{1}{2}
    \left[
      \mathbf{b}_{g}+\frac{1}{1+u^{0}}
      \left(
        \mathbf{e}_{g}\times\mathbf{u}
      \right)
    \right] +
    \frac{G_{\mathrm{F}}}{\sqrt{2}}
    \left[
      \mathbf{u}
      \left(
        g^{0}-\frac{(\mathbf{g}\mathbf{u})}{1+u^{0}}
      \right) -
      \mathbf{g}
    \right]
    \notag
    \\
    & +
    \mu
    \left[
      u^{0}\mathbf{b}-\frac{\mathbf{u}(\mathbf{u}\mathbf{b})}{1+u^{0}}+(\mathbf{e}\times\mathbf{u})
    \right]
  \bigg\}.
\end{align}
Here $\mathbf{e}_{g}$ and $\mathbf{b}_{g}$ are the components of
the tensor $G_{ab}$: $G_{ab}=(\mathbf{e}_{g},\mathbf{b}_{g})$.

The neutrino characteristics depend on the particle helicity, i.e.
the projection of the particle spin on the neutrino velocity: $h=(\bm{\zeta}\cdot\mathbf{u})/|\mathbf{u}|$.
Hence, despite accounting for the neutrino spin evolution in eq.~(\ref{eq:nuspinrot}),
we have to track the evolution of $\mathbf{u}$ in the neutrino scattering~\cite{PomKhr98},
$\tfrac{\mathrm{d}u^{a}}{\mathrm{d}t}=\tfrac{1}{U^{t}}G^{ab}u_{b}$.
If a particle moves in the equatorial plane of BH, then the asymptotic
values of $\mathbf{u}$ are~\cite{Dvo20a}, $\mathbf{u}_{\pm\infty}=\left(\pm\sqrt{E^{2}-m^{2}}/m,0,0\right)$,
where `$+$' stays for an outgoing neutrino and `$-$' for an incoming
one, $E$ is the neutrino energy, which is the integral of motion
in the metric in eq.~\eqref{eq:Kerrmetr}, and $m$ is the neutrino
mass.

Instead of solving eqs.~(\ref{eq:nuspinrot}) and~(\ref{eq:vectG}),
we can consider the evolution of the effective spinor $\psi^{\mathrm{T}}=(\psi_{1},\psi_{2})$,
with the components being responsible for the particular neutrino
polarizations. It obeys the effective Schr\"odinger equation. The coordinate
$r$, or its dimensionless analogue $x=r/r_{g}$, is more convenient
than $t$ in describing the neutrino evolution in the particle scattering
off BH. Using the results of refs.~\cite{Dvo20a,Dvo20b}, we get
that
\begin{equation}\label{eq:Schreq}
  \mathrm{i}\frac{\mathrm{d}\psi}{\mathrm{d}x}=\hat{H}_{x}\psi,
  \quad
  \hat{H}_{x}=-\mathcal{U}_{2}(\bm{\bm{\sigma}}\cdot\bm{\bm{\Omega}}_{x})\mathcal{U}_{2}^{\dagger},
\end{equation}
where $\bm{\bm{\sigma}}=(\sigma_{1},\sigma_{2},\sigma_{3})$ are the
Pauli matrices, $\bm{\bm{\Omega}}_{x}=r_{g}\bm{\bm{\Omega}}\tfrac{\mathrm{d}t}{\mathrm{d}r}$,
and $\mathcal{U}_{2}=\exp(\mathrm{i}\pi\sigma_{2}/4)$. One should
account for the matrix $\mathcal{U}_{2}$ in $\hat{H}_{x}$ since
the spin is quantized along the first\footnote{We recall that only $\mathbf{u}_{\pm\infty,1}\neq0$. Thus, the neutrino
spin should be projected on $x^{a=1}$ axis asymptotically at $r\to\infty$.} rather than the third axis.

Equation~(\ref{eq:Schreq}) should be supplied with the initial condition.
If incoming neutrinos are left polarized, i.e. $h_{-\infty}=-1$ at
$t\to-\infty$, then $\psi_{-\infty}^{\mathrm{T}}=(1,0)$. The transition
$P_{\mathrm{LR}}$ (i.e., when scattered neutrinos are right polarized
with $h_{+\infty}=+1$) and survival $P_{\mathrm{LL}}$ probabilities
in the wake of the scattering can be found on the basis of the asymptotic
solution of eq.~(\ref{eq:Schreq}), $\psi_{+\infty}^{\mathrm{T}}=(\psi_{+\infty,1},\psi_{+\infty,2})$
at $t\to+\infty$. They are\footnote{Note that the neutrino velocity changes its direction in the scattering:
$\mathbf{u}_{+\infty}=-\mathbf{u}_{-\infty}$.} $P_{\mathrm{LR}}=|\psi_{+\infty,1}|^{2}$ and $P_{\mathrm{LL}}=|\psi_{+\infty,2}|^{2}$.

In the following, we consider only utrarelativistic neutrinos moving
in the equatorial plane of BH. In general, there are nonrelativistic
cosmic neutrinos (see, e.g., ref.~\cite{RinWon04}). However, the
study of spin oscillations of such particles in questionable since
one can hardly form an initial beam of left polarized nonrelativistic
neutrinos in natural conditions.

We suppose that BH is embedded in the electromagnetic field which
asymptotically, at $r\to\infty$, approaches to the constant magnetic
field parallel to the BH rotation axis. Realistic electromagnetic
fields in the vicinity of BHs can have a complicated structure. The magnetic field in a disk has both poloidal and toroidal components. The poloidal field is generated mainly by the plasma circular motion in an accretion disk. The toroidal magnetic field results from the differential rotation of a disk. The toroidal component was mentioned in ref.~\cite{Bes10} to be dominant in many cases. However, the main contribution to the neutrino spin-flip is when a particle at the minimal distance to BH: $x=x_m$. In these moments, the toroidal component is along the neutrino velocity and does not  change the neutrino polarization. Therefore, in our work, we consider the approximation when only poloidal magnetic field is present. An electromagnetic
field in such a system was studied in ref.~\cite{Wal74}. The details
of the derivation of $F_{\mu\nu}$ are present in appendix~\ref{sec:EMBH}.

Using eqs.~(\ref{eq:Kerrmetr})-(\ref{eq:vierbKerr}), (\ref{eq:Uftr}), \eqref{eq:vectG}, and~(\ref{eq:Atphi}), we can write down the components of $\bm{\bm{\Omega}}_x$
in eq.~\eqref{eq:Schreq} as
\begin{align}\label{eq:Omegax}
  \Omega_{x1}= &
  \frac{V_{m}x^{3/2}\sqrt{x^{3}+z^{2}(x+1)}
  \left[
    \sqrt{2}x^{3/2}+\lambda(z-y)
  \right]}
  {[x^{3}+z^{2}(x+1)-zy]\sqrt{2x^{3}-3x^{2}+2\lambda\sqrt{2}zx^{3/2}}},
  \nonumber
  \\
  \Omega_{x2}= & \pm\frac{1}{2\sqrt{R(x)}}
  \bigg\{
    -\frac{V_{\mathrm{B}}}{x^{3/2}}
    \left[
      2x^{3}+z^{2}-yz
    \right]
    \notag
    \\
    & +
    \frac{y
    \left[
      x^{4}(3-2x)+xyz(3-4x)-xz^{2}(3-3x+2x^{2})-2yz^{3}+2z^{4}
    \right]}
    {2\sqrt{x(x-1)+z^{2}}\sqrt{x^{3}+z^{2}(x+1)}[x^{3}+z^{2}(x+1)-zy]}
  \bigg\},
  \nonumber
  \\
  \Omega_{x3}= & \pm\frac{V_{m}yx^{5/2}\sqrt{x(x-1)+z^{2}}
  \left[
    \sqrt{2}x^{3/2}+\lambda(z-y)
  \right]}
  {\sqrt{R(x)}[x^{3}+z^{2}(x+1)-zy]\sqrt{2x^{3}-3x^{2}+2\lambda\sqrt{2}zx^{3/2}}},
\end{align}
where $y=b/r_{g}$, $b=L/E$ is the impact parameter for ultrarelativistic
neutrinos, $L$ is the neutrino angular momentum, which is the integral
of motion, $V_{m}=G_{\mathrm{F}}n_{e}r_{g}/\sqrt{2}$ and $V_{\mathrm{B}}=\mu Br_{g}$
are the dimensionless effective potentials of the neutrino interaction
with matter of an accretion disk and with a magnetic field, and $R(x)=x^{3}+(z^{2}-y^{2})x+(y-z)^{2}$
is the dimensionless effective potential, which defines the range
of $x=r/r_{g}$ variation in the neutrino scattering. The maximal
root $x_{m}$ of the equation $R(x)=0$ gives the minimal distance
which a neutrino approaches to BH. In eq.~(\ref{eq:Omegax}), the
`$-$' sign correspond to an incoming neutrino and `$+$' one for
an outgoing one.

The derivation of eq.~\eqref{eq:Omegax} is straightforward but quite
lengthy. It is performed in refs.~\cite{Dvo13,Dvo20b}. Therefore
we omit the details here.

\section{Astrophysical applications\label{sec:APPL}}

In this section, we choose the parameters of BH, its accretion disk,
the magnetic field, and a neutrino which correspond to a realistic
situation. These parameters are used in the numerical solution of
eqs.~(\ref{eq:Schreq}) and~(\ref{eq:Omegax}).

Before we study a realistic case of BH surrounded by
a magnetized accretion disk, we should consider the situation when a neutrino
interacts only with the gravitational field of BH. For this purpose,
we set $V_{m,\mathrm{B}}=0$ in eq.~(\ref{eq:Omegax}). In this situation,
eq.~(\ref{eq:Schreq}) can be solved exactly. The probability of
transitions $L\to R$ in the gravitational neutrino scattering is
$P_{\mathrm{LR}}=(1+\cos\alpha_{+\infty}^{(g)})/2$, where
\begin{align}\label{eq:alpha+inf}
  \alpha_{+\infty}^{(g)}= &
  4\int_{x_{m}}^{\infty}\mathrm{d}x|\Omega_{x2}|
  \notag
  \\
  & =
  y\int_{x_{m}}^{\infty}\mathrm{d}x\frac{x^{4}(3-2x)+xyz(3-4x)-xz^{2}(3-3x+2x^{2})-2yz^{3}+2z^{4}}
  {\sqrt{R(x)[x(x-1)+z^{2}][x^{3}+z^{2}(x+1)]}[x^{3}+z^{2}(x+1)-zy]},
\end{align}
It turns out that $\alpha_{+\infty}^{(g)}=-\pi$ in eq.~(\ref{eq:alpha+inf})
for any $z\leq1/2$ and $y>y_{0}=4\cos^{3}\left[\tfrac{1}{3}\arccos(\mp2z)\right]\pm z$.
Double signs in the expression for the critical impact parameter $y_{0}$,\footnote{If $y \leq y_{0}$, an ultrarelativistic particle falls into BH~\cite{BarPreTeu72}.}
correspond to the positive or negative projection of the neutrino
angular momentum to the spin of BH.

The fact that $\alpha_{+\infty}^{(g)}=-\pi$ in the purely gravitational
neutrino scattering means that $P_{\mathrm{LR}}=0$, i.e. there is
no spin oscillations of ultrarelativistic neutrinos. This result is
in agreement with the finding of ref.~\cite{Lam05}, where the same
result was obtained in the weak field limit and considering the chiral
neutrino eigenstates as incoming and outgoing wave functions. Note
that the concepts of the helicity and chirality are not the same for
an ultrarelativistic fermion in curved spacetime, as established in
ref.~\cite{Sin05}. We have proven the absence of spin oscillations
of ultrarelativistic neutrinos meaning the transitions between the
helicity states $(\bm{\bm{\zeta}}\cdot\mathbf{u})/|\mathbf{u}| = \pm 1$.
Moreover, this fact holds true for the arbitrary particle motion in
the allowed region in the gravitational scattering in the Kerr metric.
This our finding corrects the statement of ref.~\cite{Dvo20b}, where
$P_{\mathrm{LR}}\neq0$ was obtained for ultrarelativistic neutrinos
scattered off a rotating BH.

The inclusion of the term $\propto V_{\mathrm{B}}$ to $\Omega_{x2}$
in eq.~(\ref{eq:Omegax}) makes the neutrino spin to precess even
in the ultrarelativistic case. However, the approximation when $V_{\mathrm{B}}=\text{const}$
in eq.~(\ref{eq:Omegax}) leads to the divergent value of $\alpha_{+\infty}$.
It results from the fact that the magnetic field at $r\to\infty$
is uniform and nonzero. Thus the neutrino spin makes an infinite number of revolutions
with respect to its initial direction even when a particle is far
away from BH. Therefore, the approximation $V_{\mathrm{B}}=\text{const}$
is unphysical and we have to replace $B=\text{const}$ with some decreasing
function of $x$.

As a rule, the electron number density $n_{e}$ also decreases with
the radius. If we consider the neutrino scattering off SMBH with $M\sim10^{8}\,M_{\odot}$,
then $n_{e}(r)\propto n_{e}^{(0)}r^{-\beta}$, where the number density
in the vicinity of SMBH is $n_{e}^{(0)}\sim10^{18}\,\text{cm}^{-3}$~\cite{Jia19}. Such a dense matter can exist in the inner part of an accretion disk in some active galactic nuclei (AGN).
The index $\beta>0$ is quite model dependent. We take $\beta=3/2$,
which corresponds to an advection dominated accretion flow~\cite{NarYi94}.

The structure of the electromagnetic field in an accretion disk is
quite complicated. Some models are reviewed in ref.~\cite{Bes10}.
We can assume that $B=B_{0}x^{-\kappa}$, where $\kappa>1$. It corresponds
to the magnetic field decreasing towards the edge of an accretion
disk. The indexes $\kappa$ and $\beta$ are related by the formula,
$2\kappa=\beta+1$. Indeed, if we assume the equipartition of the
energy of the magnetic field and the plasma accreting to BH~\cite{BlaPay82},
$B^{2}\propto nv^{2}$, one gets the required relation between $\beta$
and $\kappa$. Here $v$ is the velocity of plasma which scales as
the Keplerian one, i.e. as $v\propto r^{-1/2}$. For $\beta=3/2$,
chosen above, one gets that $\kappa=5/4$~\cite{BlaPay82}. We use
this scaling law for the magnetic field.

The magnetic field $B_{0}$ in the vicinity of BH, $x\sim1$, is constrained
by the Eddington limit~\cite[pg.~184]{Bes10}, $B_{\mathrm{Edd}}=10^{4}\,\text{G}\left(M/10^{9}\,M_{\odot}\right)^{-1/2}$,
which is the upper bound. Typically, $B_{0}\ll B_{\mathrm{Edd}}$.
Then, we adopt $B_{0}=10^{-2}B_{\mathrm{Edd}}$ in our analysis. For
$M\sim10^{8}\,M_{\odot}$, it gives one $B_{0}=3.2\times10^{2}\,\text{G}$. Such magnetic fields are reported in ref.~\cite{Dal19} to exist in the vicinity of some AGN. Magnetic fields near Sgr A* or M87 are weaker (see, e.g., ref.~\cite{Eat13}).

We take the most conservative value for the neutrino magnetic moment
$\mu=10^{-14}\,\mu_{\mathrm{B}}$, where $\mu_{\mathrm{B}}$ is the
Bohr magneton, which is the model independent constraint on the Dirac
neutrino magnetic moment obtained in ref.~\cite{Bel05}. Thus, for
$M=10^{8}\,M_{\odot}$, the dimensionless interaction potentials in
eq.~(\ref{eq:Omegax}) are $V_{m}(x)=10^{-1}\times x^{-3/2}$ and $V_{\mathrm{B}}(x)=2.7\times10^{-2}\times x^{-5/4}$
since $r_{g}=2.95\times10^{13}\text{cm}=1.5\times10^{27}\text{GeV}^{-1}$.

Now, we can discuss the general case of the neutrino scattering off
a rotating BH surrounded by a magnetized accretion disk. We should
also mention that, in practice, one cannot track a particular neutrino
and measure the change of its polarization. A neutrino telescope can
detect a flux of neutrinos scattered off BH. If neutrinos were scalar
particles, the outgoing flux could be defined as $F_{0}$.\footnote{Note that $F_{0}\propto(\mathrm{d}\sigma/\mathrm{d}\varOmega)_0$, which
is the differential cross section for scalar particles. Here $\mathrm{d}\varOmega = 2\pi \sin\chi \mathrm{d}\chi$, where $\chi$ is the scattering angle.} The measured flux of spinning neutrinos is $\propto P_{\mathrm{LL}}F_{0}$
since a terrestrial detector is sensitive to left polarized ultrarelativistic
neutrinos only.

Different neutrino trajectories contribute to the flux in a detector
with a particular scattering angle $0<|\chi|<\pi$. In general relativity,
a particle can make multiple revolutions around BH before being scattered
off. In our simulations, we account for up to four such revolutions.
Moreover, the $\theta$-symmetry is broken in the Kerr metric. Thus,
despite we consider the neutrino motion in the equatorial plane only,
we can formally consider both positive and negative scattering angles
$\chi$. Positive $\chi$ correspond to a detector inclined towards
the BH rotation, whereas, when $\chi<0$, it is inclined in the opposite
direction. These two situations were explained in ref.~\cite{Dvo20b}
to result in the different scattering pictures. We call them the direct
and retrograde scatterings.

The strategy for the numerical solution of eqs.~(\ref{eq:Schreq})
and~(\ref{eq:Omegax}) is the following. First, we integrate eq.~(\ref{eq:Schreq})
from $x=\infty$ to $x=x_{m}$ with the initial condition $\psi=\psi_{-\infty}$.
Then, we account for the change of the signs in $\Omega_{x2,3}$ in
eq.~(\ref{eq:Omegax}). We use the result of the first integration
$\psi_{m}=\psi(x_{m})$ as the initial condition in the second numerical
integration from $x=x_{m}$ to $x=\infty$. Finally, we find the
spin state of outgoing neutrinos using $\psi_{+\infty}=\psi(x=\infty)$.

\section{Results\label{sec:RES}}

In this section, we present the measured fluxes of neutrinos based
on the numerical solution of eqs.~(\ref{eq:Schreq}) and~(\ref{eq:Omegax})
with the initial conditions and the parameters given in section~\ref{sec:APPL}.

First, we study the situation when only the magnetic field contribution
is nonzero in eq.~(\ref{eq:Omegax}). In this case, only $\Omega_{x2}\neq0$
in eq.~(\ref{eq:Omegax}). Thus, eq.~(\ref{eq:Schreq}) can be solved
in quadratures. We have already found in section~\ref{sec:APPL} that
solely the gravitational field does not contribute to the spin-flip
of scattered neutrinos. The angle of the neutrino spin rotation owing
to the magnetic field interaction reads
\begin{equation}
  \alpha_{+\infty}^{(\mathrm{B})}=
  2V_{\mathrm{B}}^{(0)}\int_{x_{m}}^{\infty}\mathrm{d}x\frac{2x^{3}+z^{2}-yz}{x^{11/4}\sqrt{R(x)}},
\end{equation}
where $V_{\mathrm{B}}^{(0)}=2.7\times10^{-2}$. The survival probability
is $P_{\mathrm{LL}}^{(\mathrm{B})}=(1-\cos\alpha_{+\infty}^{(\mathrm{B})})/2$.

We note that, unlike the solely gravitational interaction considered
in section~\ref{sec:APPL}, $P_{\mathrm{LL}}^{(\mathrm{B})}<1$ for
any $y>y_{0}$ and $z\leq1/2$. It means that there is a neutrino
spin-flip owing to the presence of the neutrino magnetic moment. Using
$P_{\mathrm{LL}}^{(\mathrm{B})}$, we calculate the flux of outgoing
neutrinos basing on the flux $F_{0}$ of scalar particles. The calculation
of $F_{0}$ is performed in the standard manner (see, e.g., ref.~\cite{DolDorLas06}).

\begin{figure}
  \centering
  \subfigure[]
  {\label{fig:1a}
  \includegraphics[scale=.33]{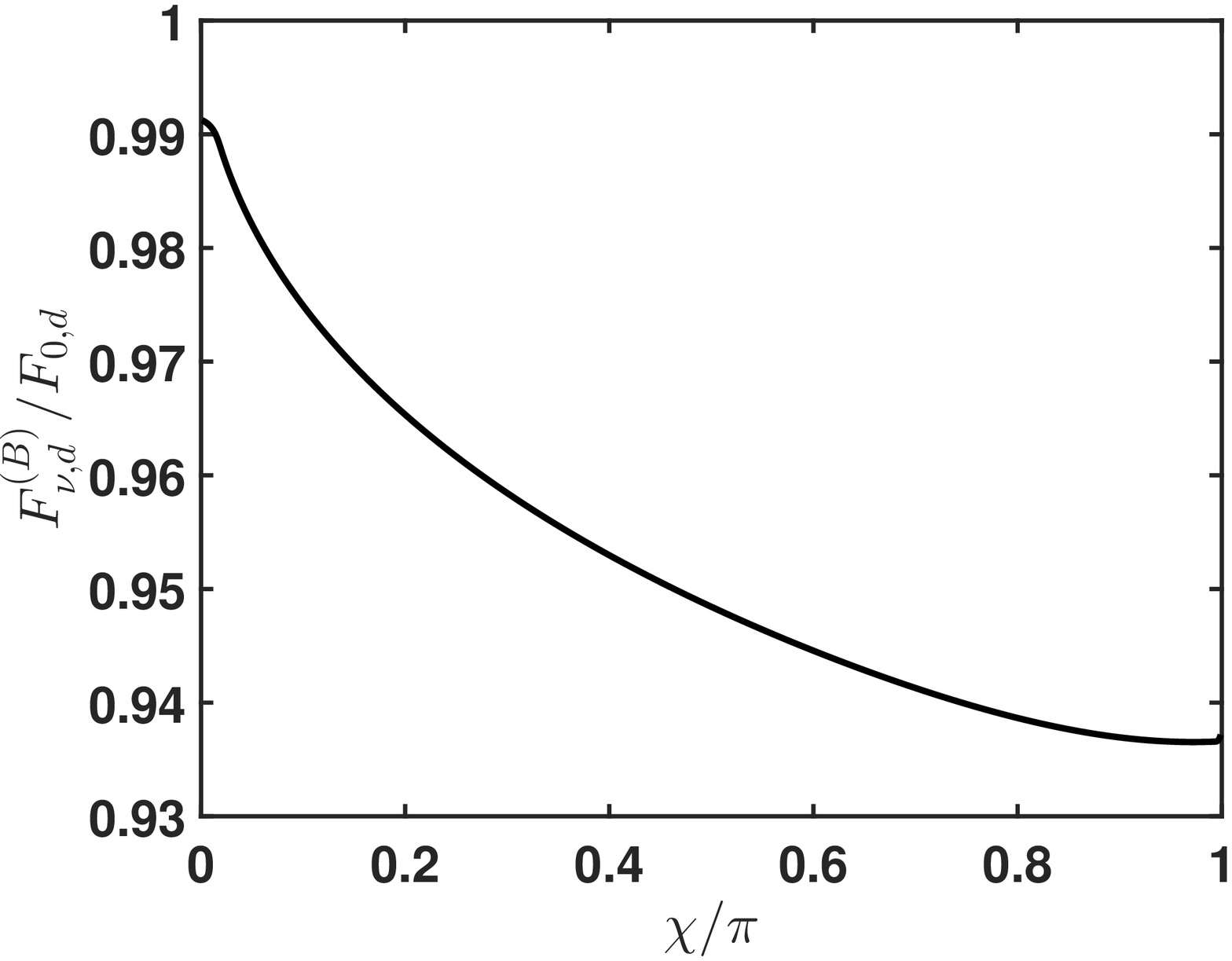}}
  \hskip-.6cm
  \subfigure[]
  {\label{fig:1b}
  \includegraphics[scale=.33]{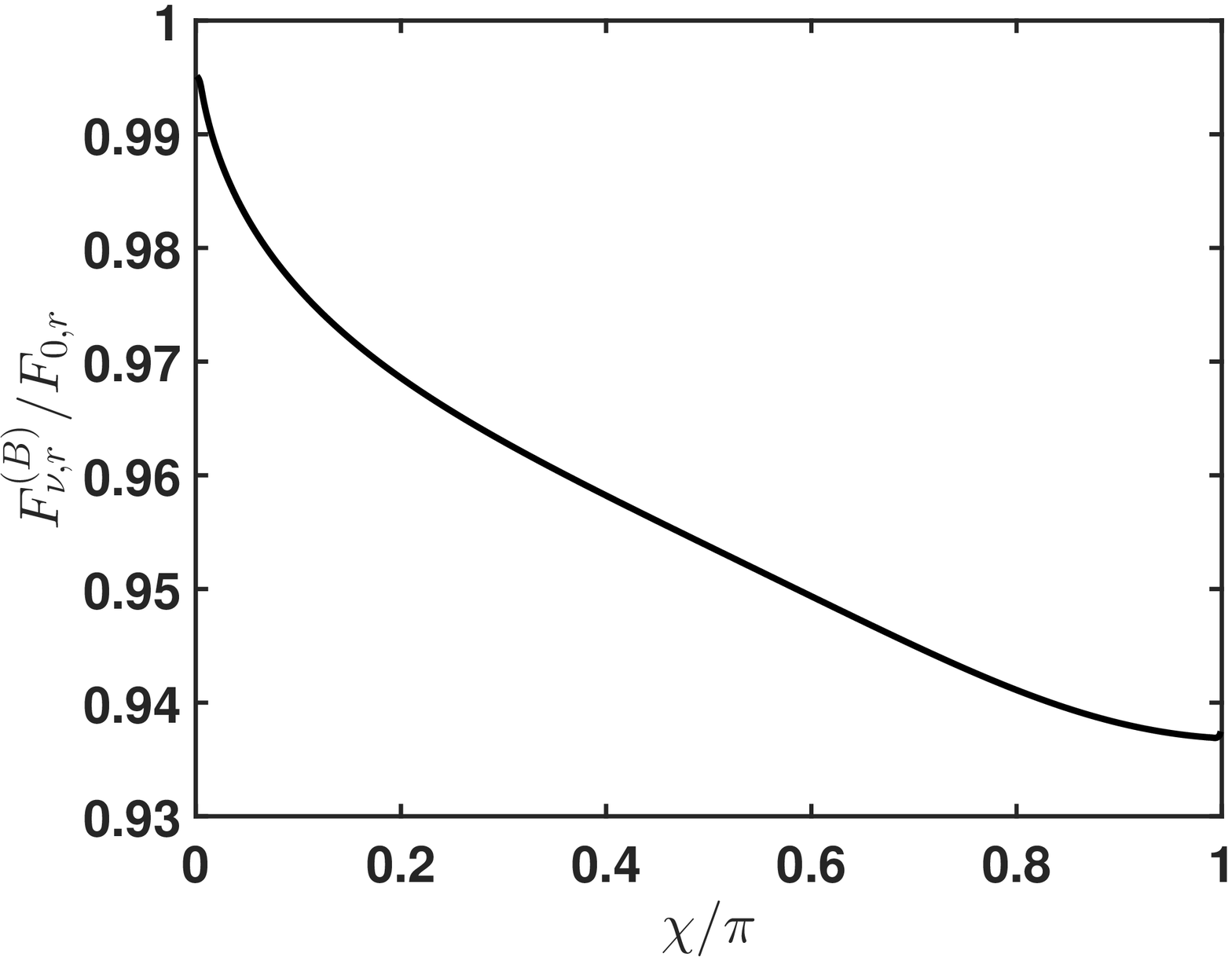}}
  \protect
  \caption{The ratios of the fluxes $F_{\nu}^{(\mathrm{B})}$ and $F_{0}$ versus
  the scattering angle $\chi$. The parameters of the system are $\mu=10^{-14}\,\mu_{\mathrm{B}}$,
  $B_{0}=3.2\times10^{2}\,\text{G}$, and $M=10^{8}\,M_{\odot}$.
  (a) Direct scattering; (b) retrograde scattering.\label{fig:magnetic}}
\end{figure}

In figure~\ref{fig:magnetic}, we show the ratio $F_{\nu,d,r}^{(\mathrm{B})}/F_{0,d,r}$
for the direct and retrograde scatterings off a maximally rotating
SMBH with $z=1/2$ versus $\chi$. One can see that this ratio monotonically
decreases from unity\footnote{The value of $F_{\nu,d,r}^{(\mathrm{B})}/F_{0,d,r}$ is slightly less
than 1 at $\chi=0$. This deviation is caused by errors of numerical
simulations.} for the forward scattering, $\chi=0$, to about 0.94 for the backward
scattering, $\chi=\pi$. There is a negligible dependence of the fluxes on the type
of the scattering, $F_{\nu,d}^{(\mathrm{B})}/F_{0,d}\approx F_{\nu,r}^{(\mathrm{B})}/F_{0,r}$.
Therefore the flux of outgoing neutrinos is reduced by $\approx(6\pm1)\%$
compared to the case when spin effects are not accounted for.

\begin{figure}
  \centering
  \subfigure[]
  {\label{fig:2a}
  \includegraphics[scale=.33]{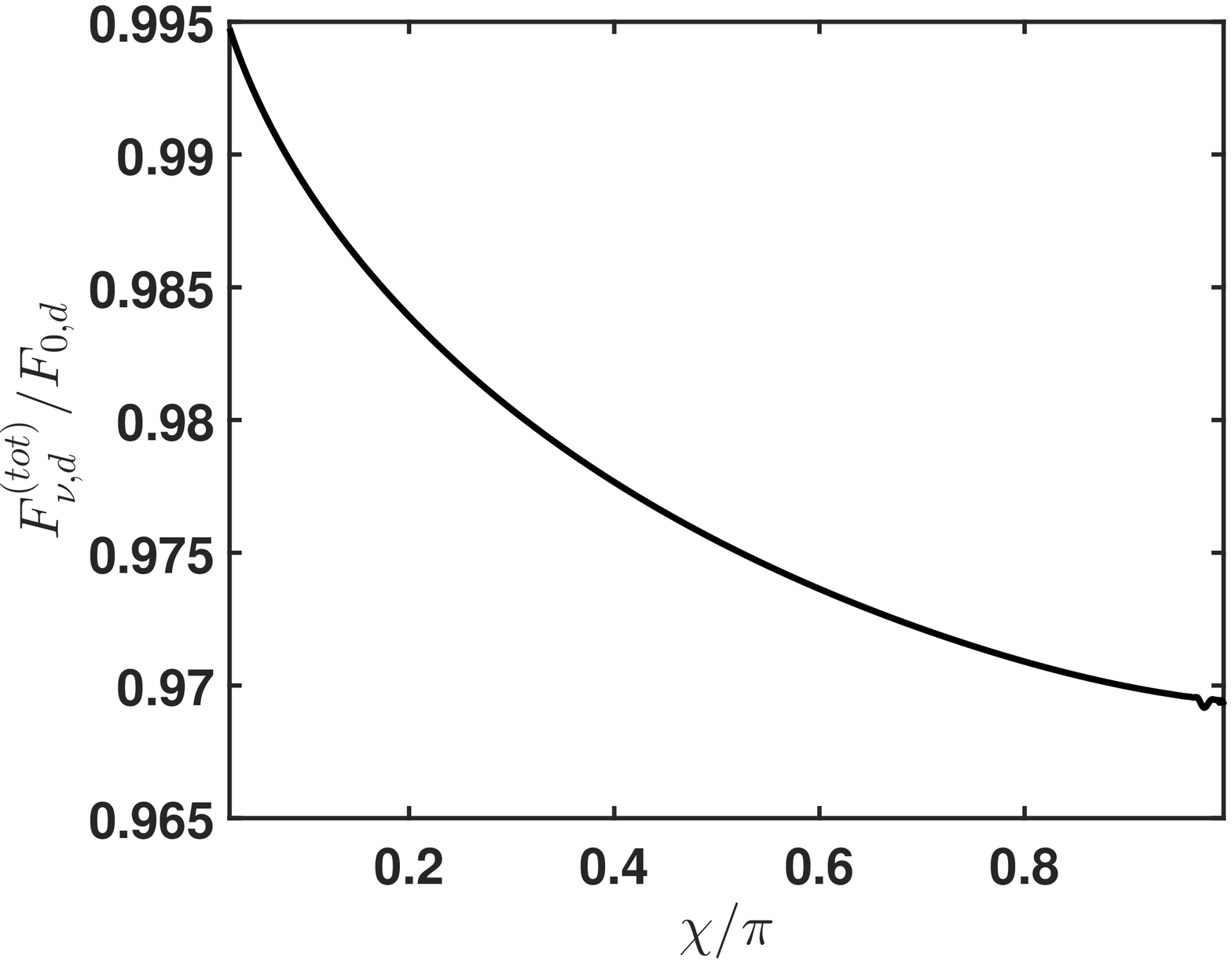}}
  \hskip-.6cm
  \subfigure[]
  {\label{fig:2b}
  \includegraphics[scale=.33]{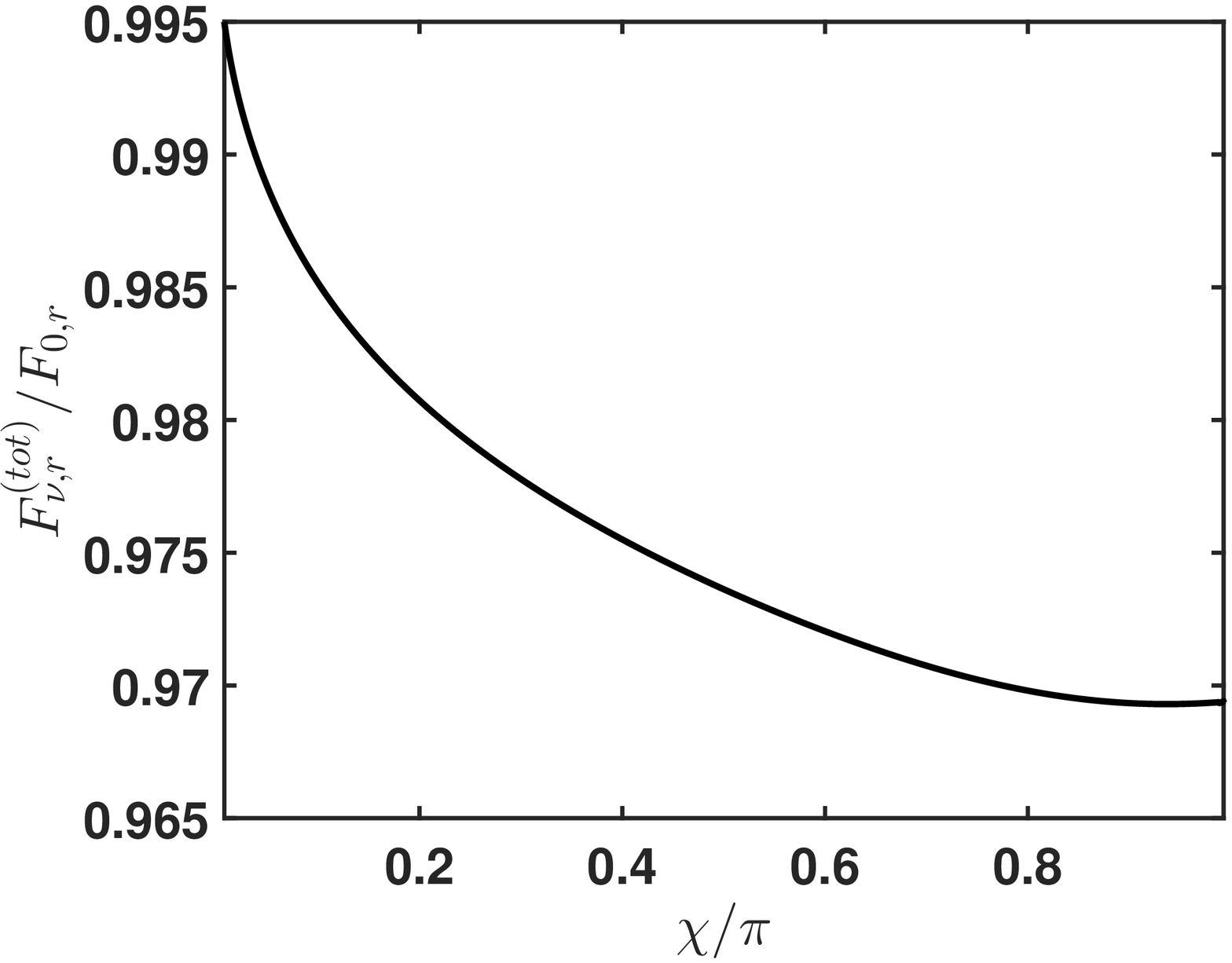}}
  \\
  \subfigure[]
  {\label{fig:2c}
  \includegraphics[scale=.33]{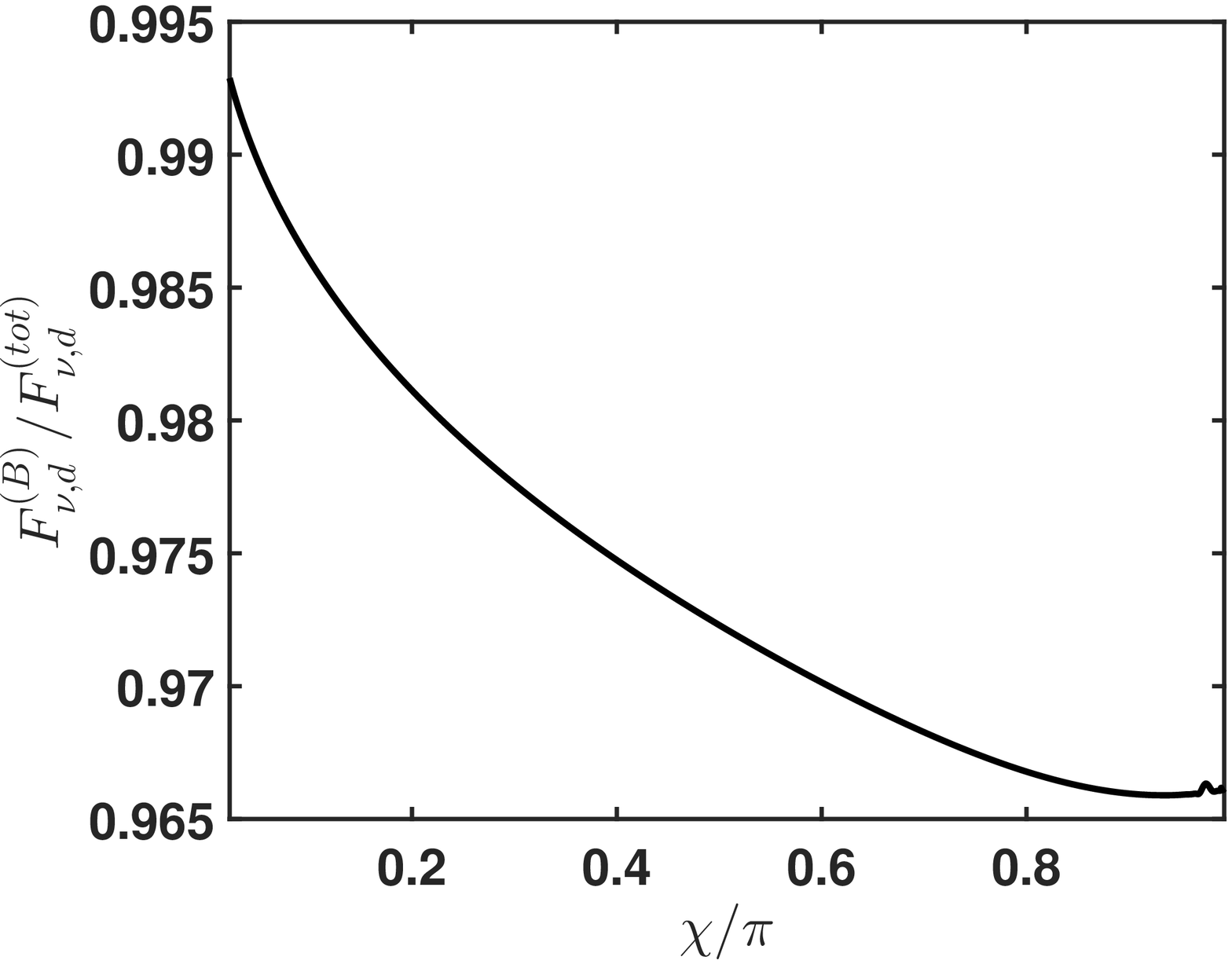}}
  \hskip-.6cm
  \subfigure[]
  {\label{fig:2d}
  \includegraphics[scale=.33]{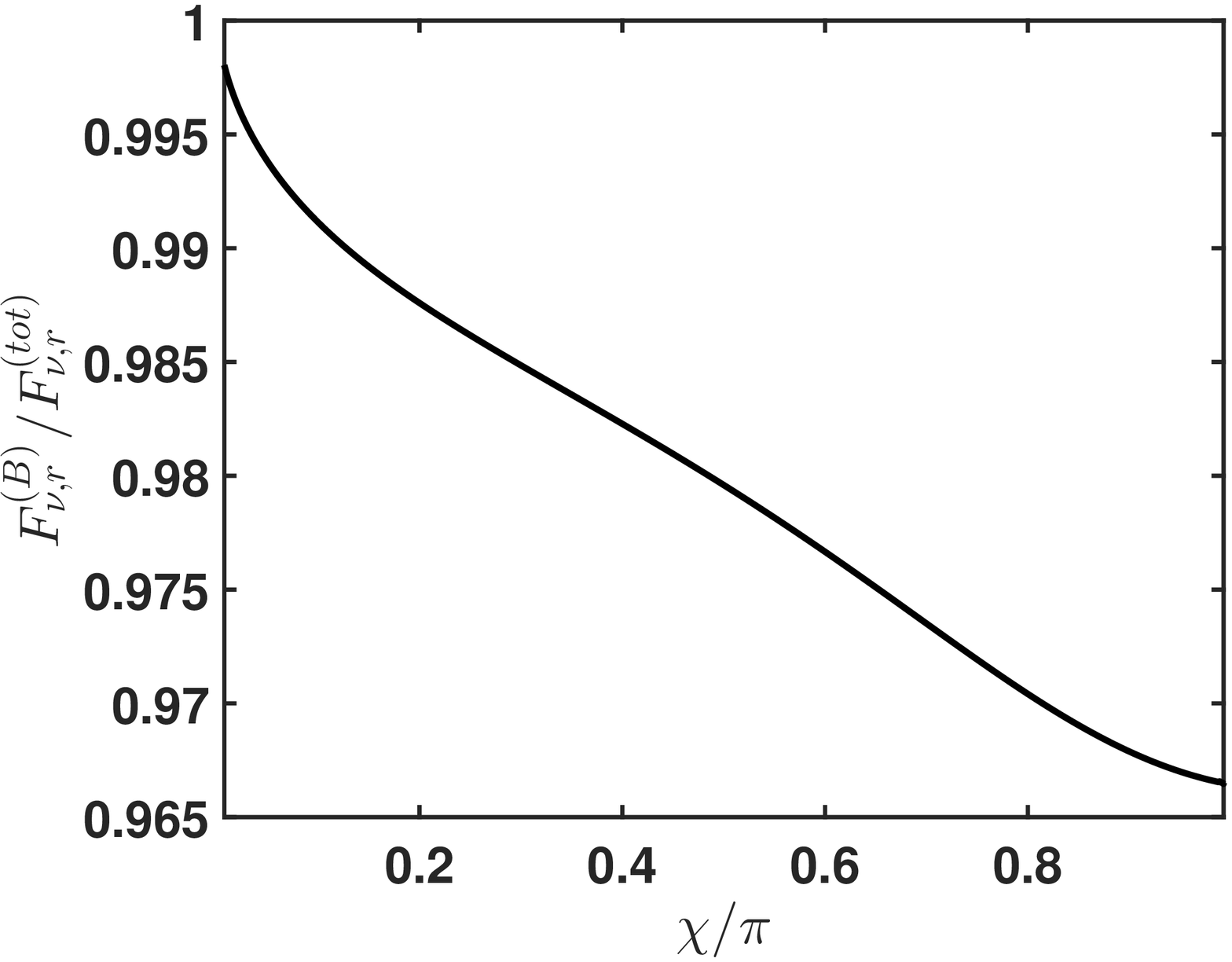}}
  \protect
  \caption{The outgoing fluxes $F_{\nu}^{(\text{tot})}$ based on the numerical
  solution of eq.~(\ref{eq:Schreq}) versus the scattering angle for
  SMBH with the maximal spin $z=1/2$ surrounded by a corotating accretion
  disk ($\lambda=+1$). The magnetic parameters are the same as in figure~\ref{fig:magnetic}
  and $n_{e}^{(0)}=10^{18}\,\text{cm}^{-3}$.
  (a) and (c) Direct scattering; (b) and (d) retrograde scattering.\label{fig:totalcorot}}
\end{figure}

Now, we take into account the neutrino interaction with an accretion disk.
In this case, all the components of $\bm{\bm{\Omega}}_{x}$ in eq.~(\ref{eq:Omegax})
are nonzero. It means that we have to solve eq.~(\ref{eq:Schreq})
numerically. We start with the situation of the disk which corotates
SMBH, i.e. $\lambda=+1$ in eq.~(\ref{eq:Omegax}). In figures~\ref{fig:2c}
and~\ref{fig:2d}, we show the ratios $F_{\nu,d,r}^{(\mathrm{B})}/F_{\nu,d,r}^{(\mathrm{tot})}$
for the direct and retrograde scatterings. One can see that $F_{\nu}^{(\mathrm{tot})}$
is $\approx3\%$ greater than $F_{\nu}^{(\mathrm{B})}$. The result that
$F_{\nu}^{(\mathrm{tot})}>F_{\nu}^{(\mathrm{B})}$ stems from the
fact that the nonzero components $\Omega_{x1,3}\neq0$ withdraw the
system from the resonance condition implemented in the case when only
$\Omega_{x2}\neq0$.

In figures~\ref{fig:2a} and~\ref{fig:2b},
we show the ratios $F_{\nu,d,r}^{(\mathrm{tot})}/F_{0,d,r}$ for the
direct and retrograde scatterings. One can see that this ratio is
minimal at $\chi=\pi$. The fluxes $F_{\nu}^{(\mathrm{tot})}(\chi=\pi)$
are reduced $\approx3\%$ compared to the scalar particles case. This
value can be explained for if we account for figures~\ref{fig:magnetic},
\ref{fig:2c}, and~\ref{fig:2d}.

Figure~\ref{fig:totalcorot} corresponds to $z=1/2$. We analyzed
the situations $z<1/2$. The behavior of the neutrino fluxes $F_{\nu,d,r}$
does not change qualitatively.

\begin{figure}
  \centering
  \subfigure[]
  {\label{fig:3a}
  \includegraphics[scale=.33]{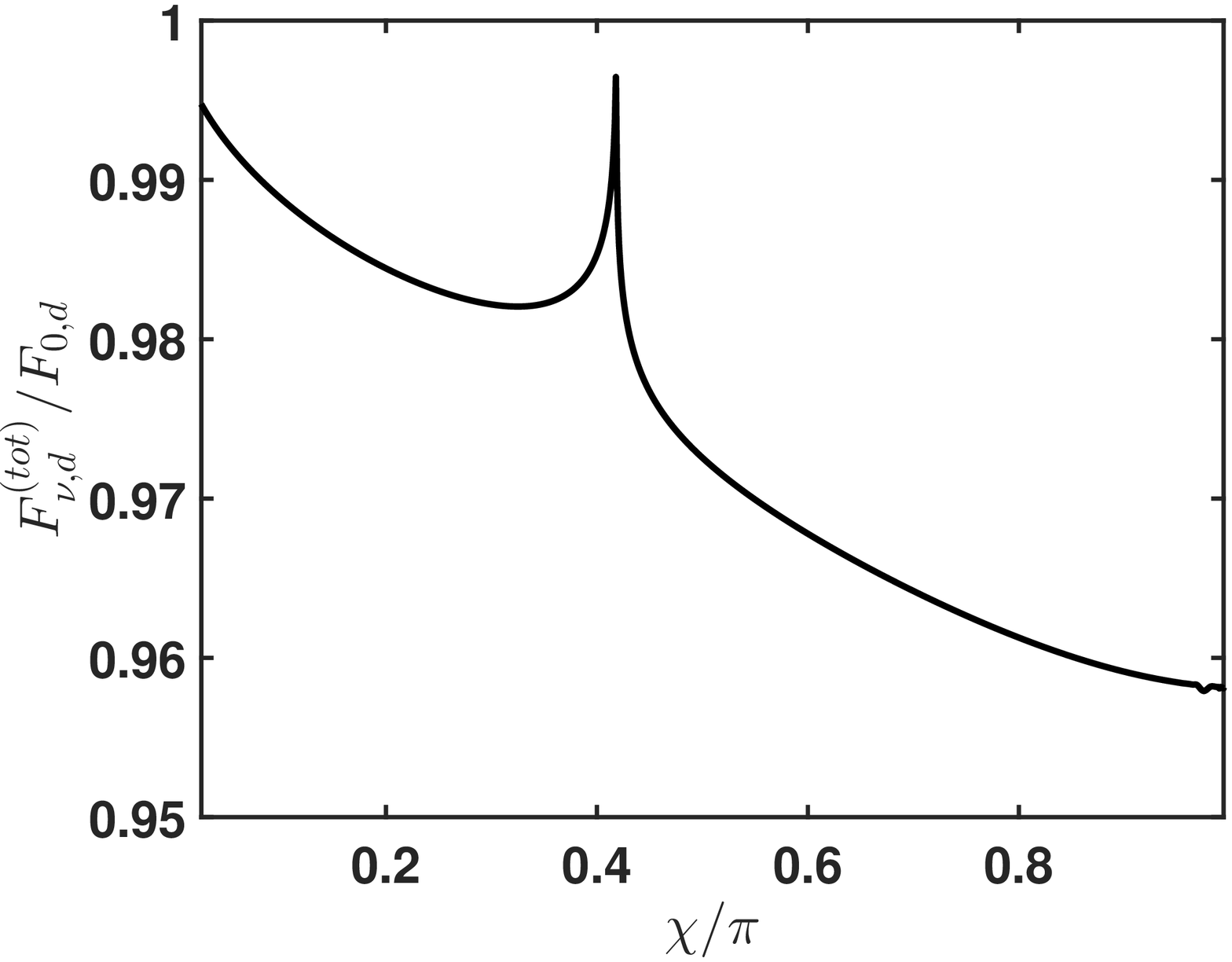}}
  \hskip-.6cm
  \subfigure[]
  {\label{fig:3b}
  \includegraphics[scale=.33]{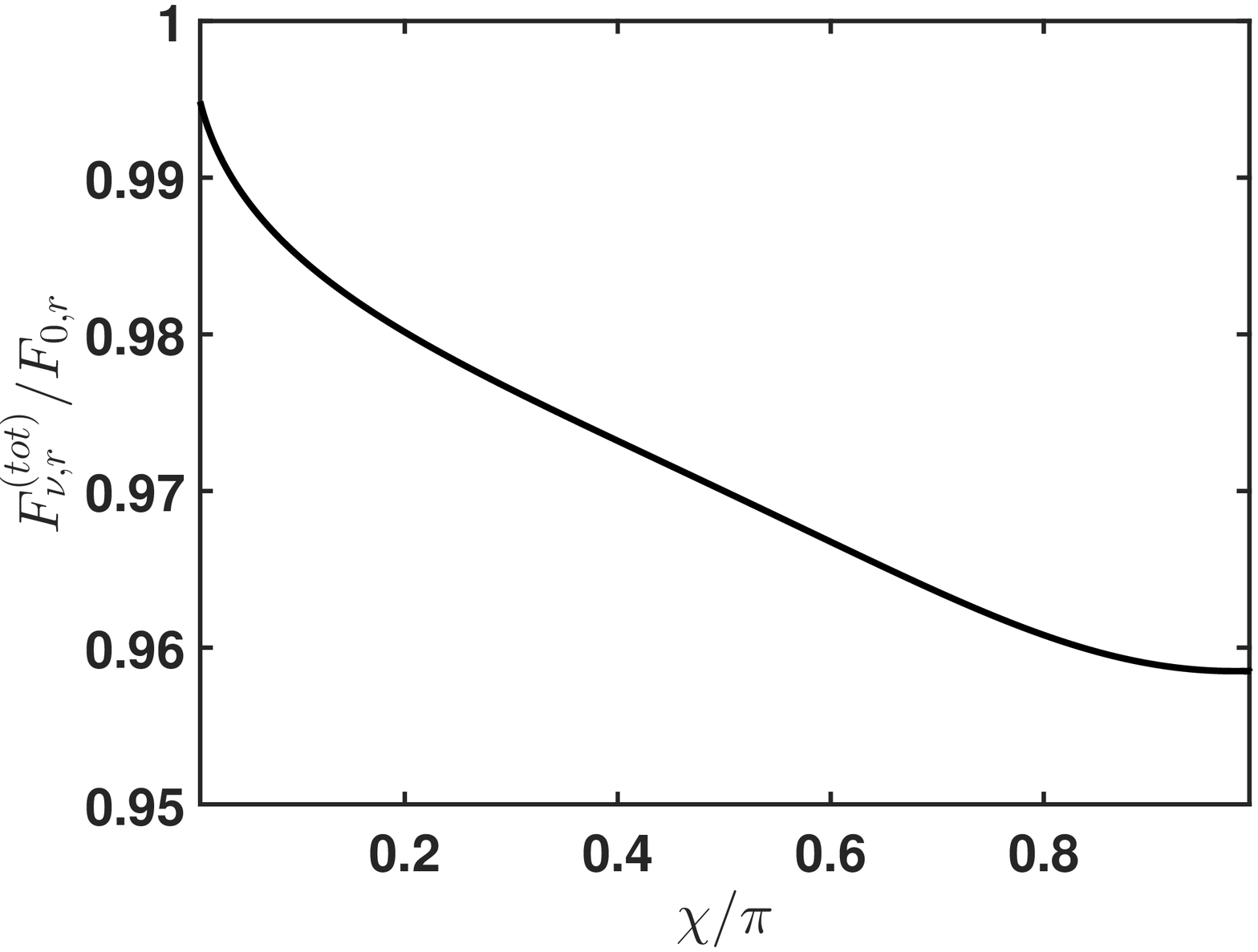}}
  \\
  \subfigure[]
  {\label{fig:3c}
  \includegraphics[scale=.33]{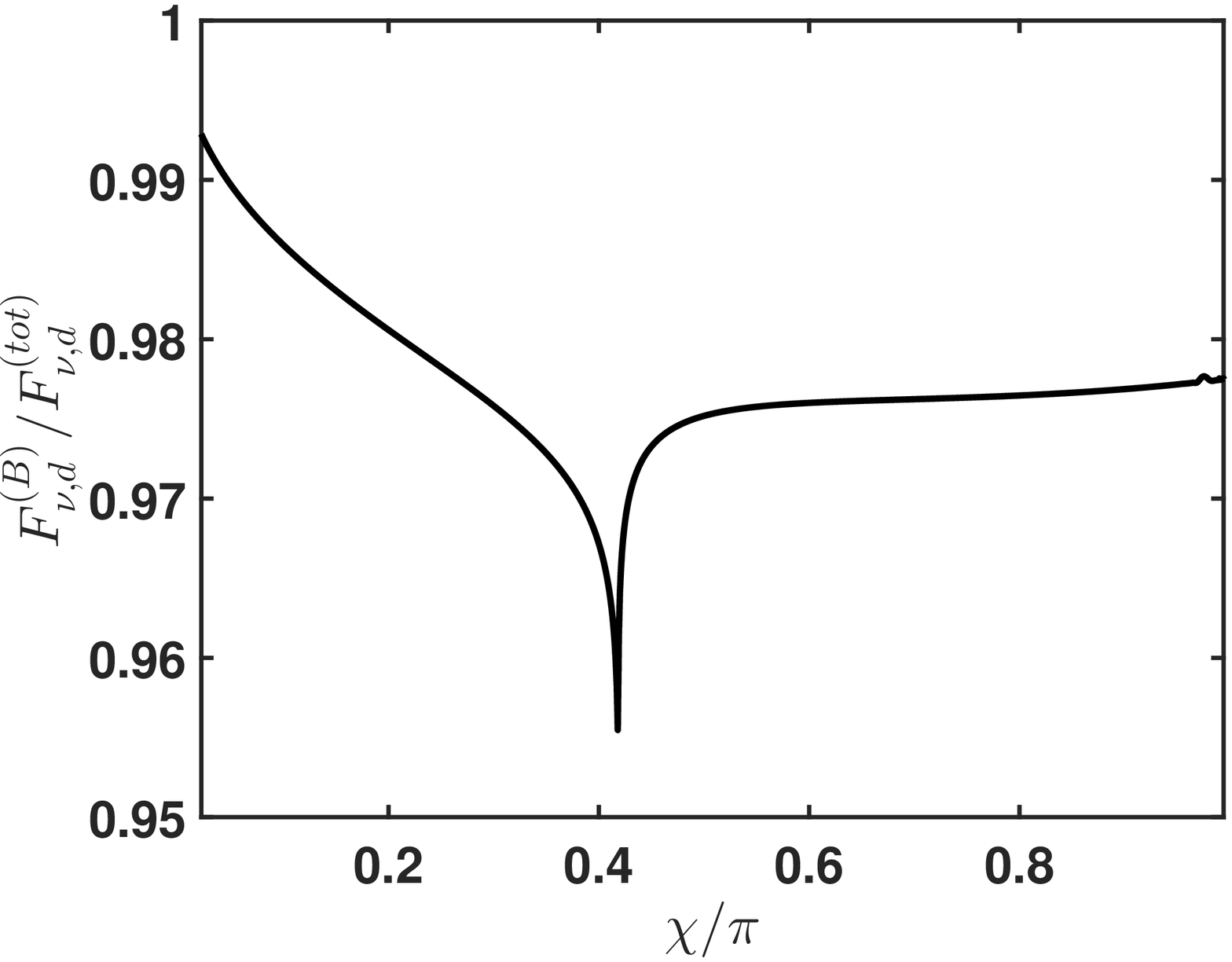}}
  \hskip-.6cm
  \subfigure[]
  {\label{fig:3d}
  \includegraphics[scale=.33]{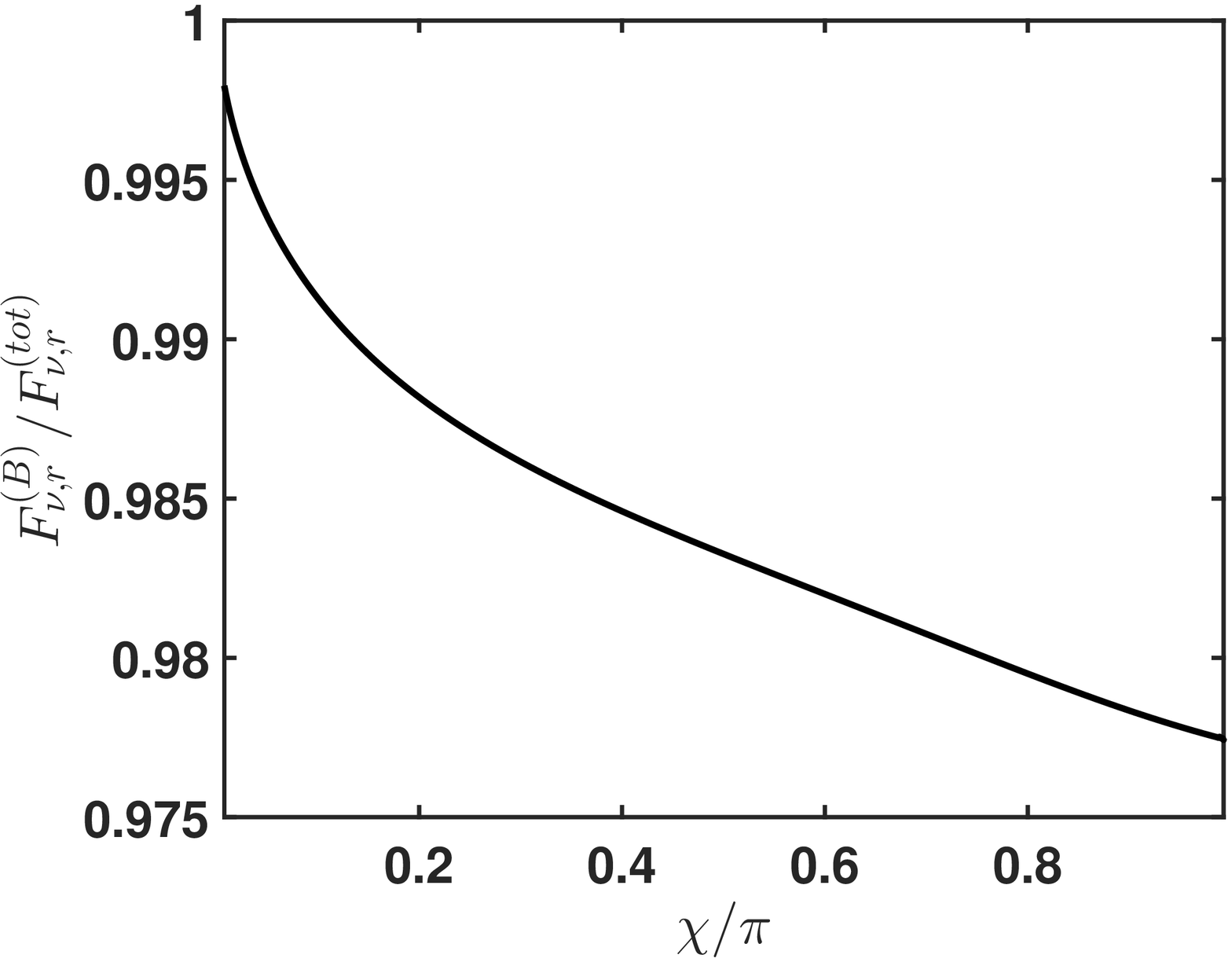}}
  \protect
  \caption{The outgoing fluxes $F_{\nu}^{(\text{tot})}$ based on the numerical solution of eq.~(\ref{eq:Schreq}) versus the scattering angle for 
  SMBH with maximal spin $z=1/2$ surrounded by a counter-rotorating accretion disk ($\lambda=-1$). The parameters of the system are the
  same as in figure~\ref{fig:totalcorot}.
  (a) and (c)~Direct scattering; (b) and (d) retrograde scattering.\label{fig:totalcounter05}}
\end{figure}

Now, we turn to an accretion disk which counter-rotates SMBH, i.e.
$\lambda=-1$. The parameters of the disk and the neutrino are the
same as in figure~\ref{fig:totalcorot}. The flux of scattered neutrinos,
obtained from the numerical solution of eq.~(\ref{eq:Schreq}), for
$z=1/2$ and $\lambda=-1$ is shown in figure~\ref{fig:totalcounter05}.
The retrograde scattering, presented in figures~\ref{fig:3b}
and~\ref{fig:3d}, does not differ qualitatively
from the case $\lambda=+1$, shown in figures~\ref{fig:2b}
and~\ref{fig:2d}.

The direct scattering, depicted figures~\ref{fig:3a} and~\ref{fig:3c},
on its turn, reveals the spike in the total flux. This feature happens
when the total survival probability, accounting for both matter and
the magnetic field, is great, $P_{\mathrm{LL}}^{(\mathrm{tot})}\approx1$.
It takes place at a certain impact parameter. 

\begin{figure}
  \centering
  \subfigure[]
  {\label{fig:4a}
  \includegraphics[scale=.33]{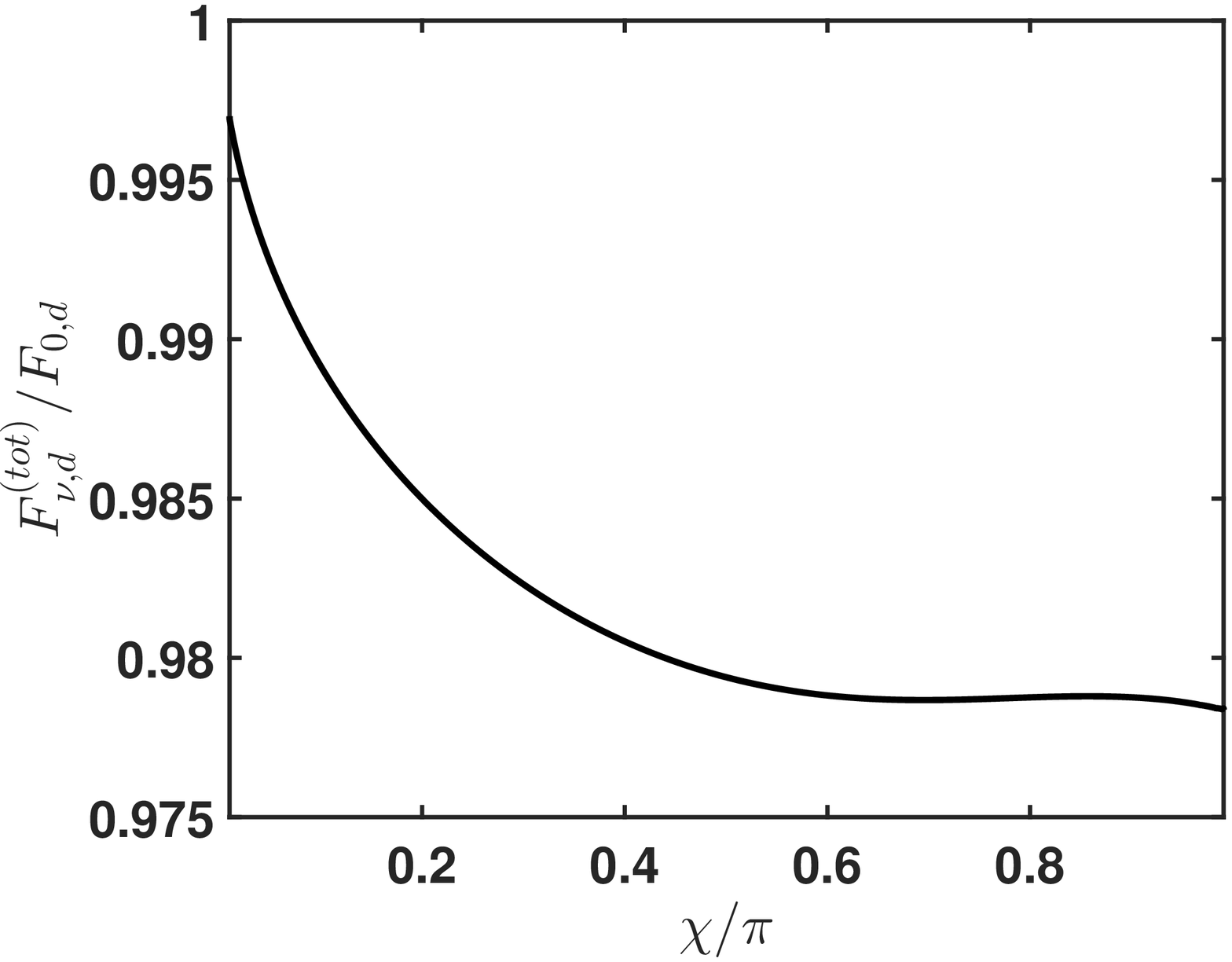}}
  \hskip-.6cm
  \subfigure[]
  {\label{fig:4b}
  \includegraphics[scale=.33]{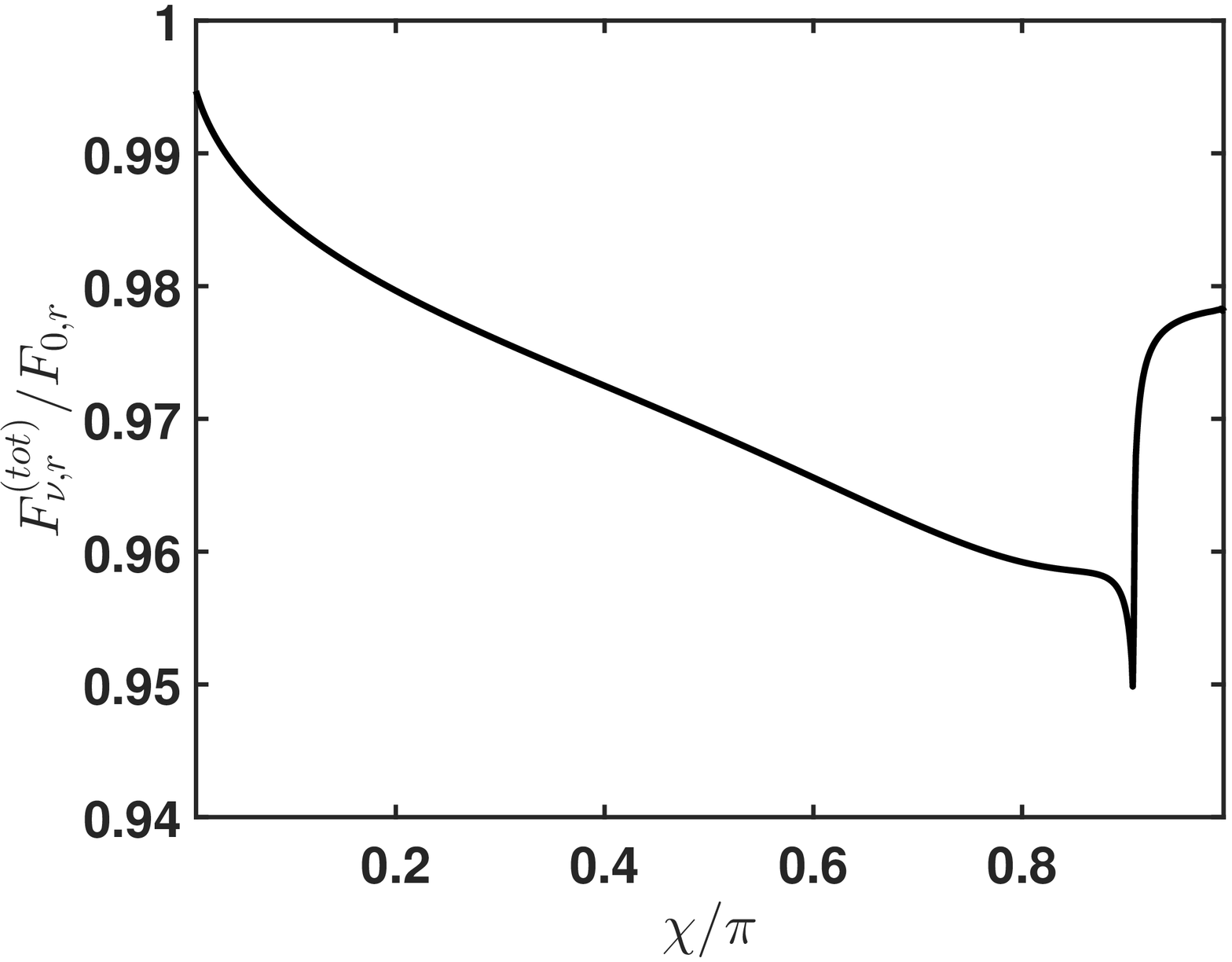}}
  \\
  \subfigure[]
  {\label{fig:4c}
  \includegraphics[scale=.33]{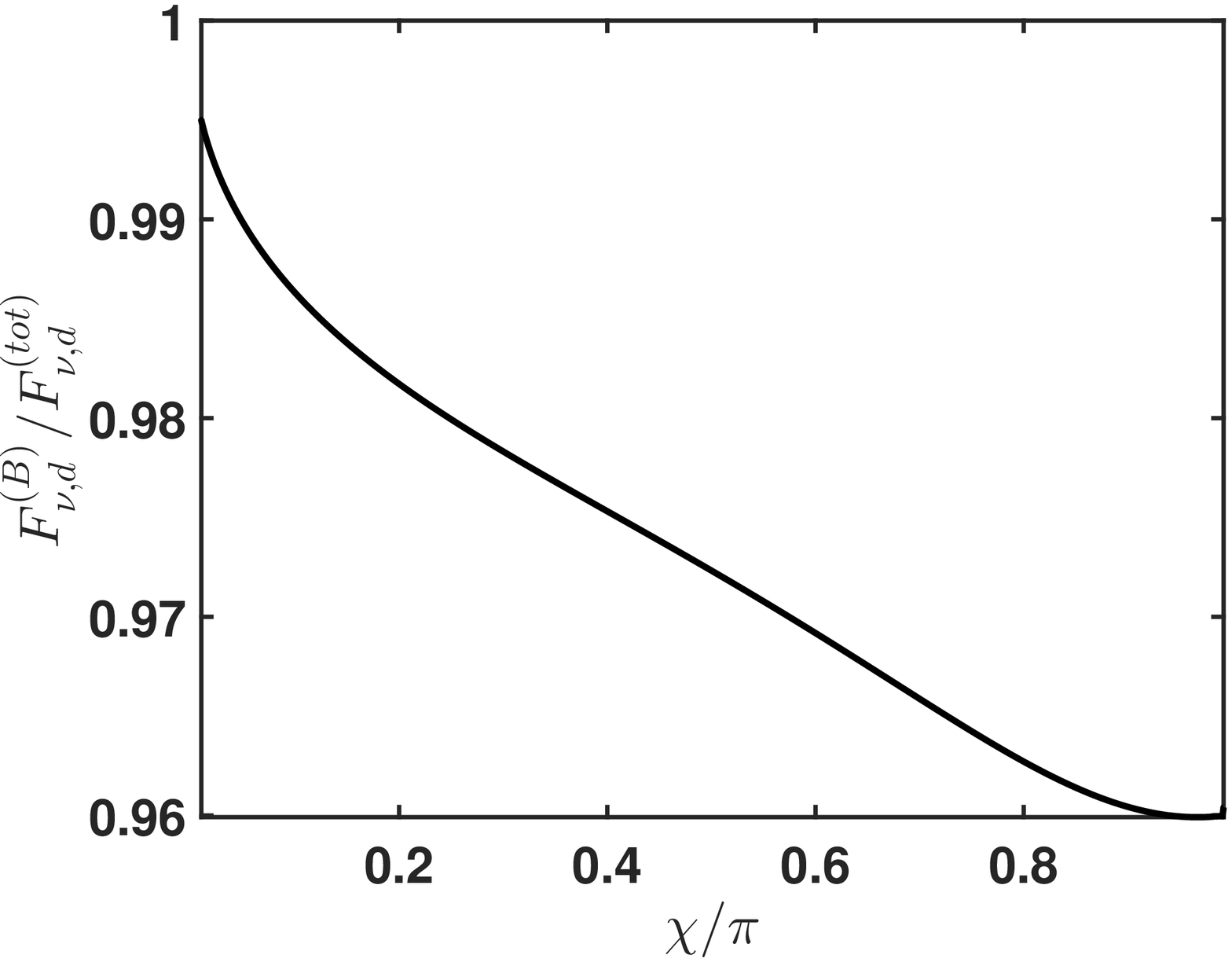}}
  \hskip-.6cm
  \subfigure[]
  {\label{fig:4d}
  \includegraphics[scale=.33]{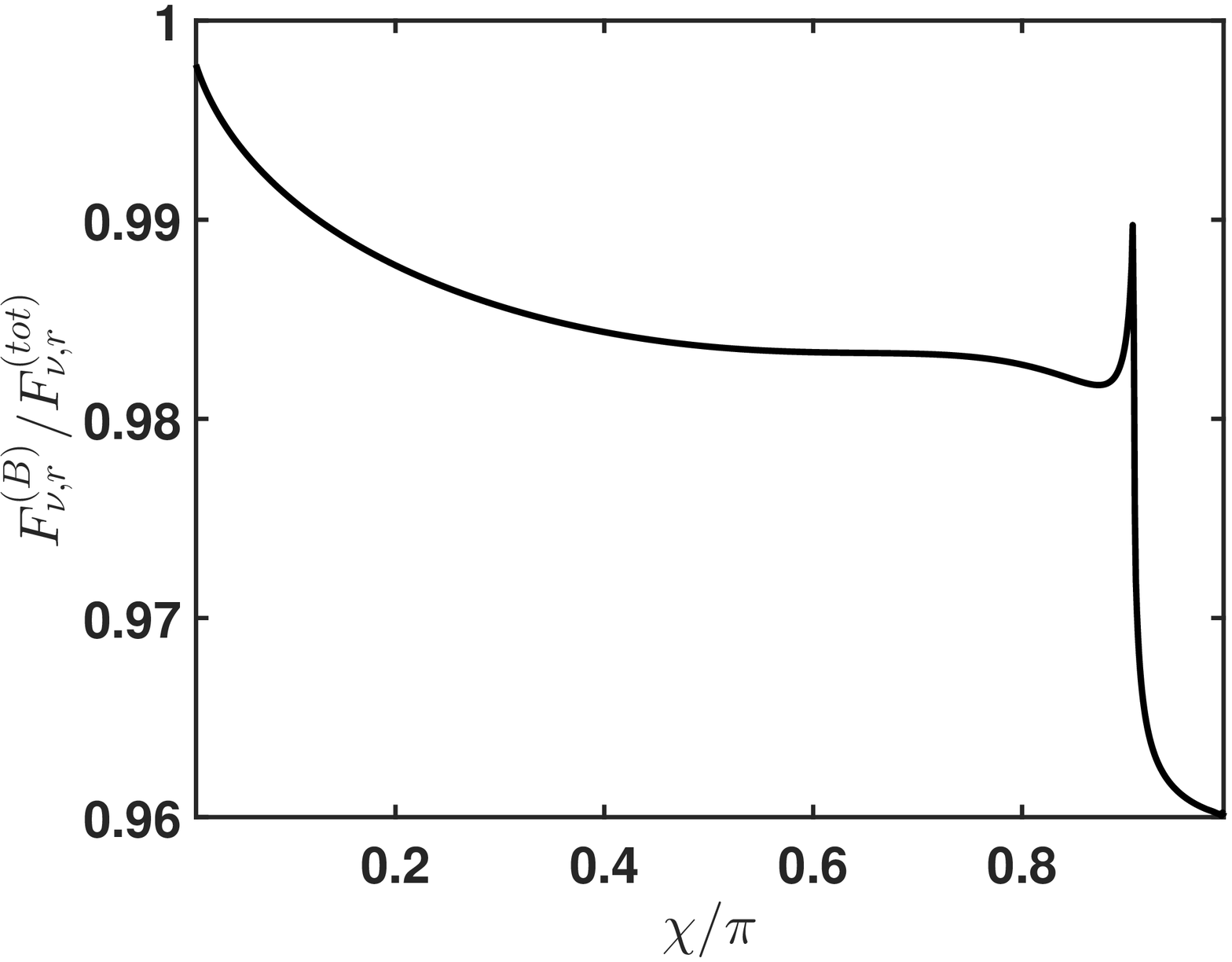}}
  \protect
  \caption{The outgoing fluxes $F_{\nu}^{(\text{tot})}$ based on the numerical solution of eq.~(\ref{eq:Schreq}) versus the scattering angle for
  SMBH surrounded by a counter-rotorating accretion disk ($\lambda=-1$). The parameters of the system are the same as
  in figure~\ref{fig:totalcorot} except for $z=0.1$.
  (a) and (c) Direct scattering; (b) and (d) retrograde scattering.\label{fig:totalcounter01}}
\end{figure}

We can see the analogous behavior of $F_{\nu}^{(\mathrm{tot})}$ for
$z=0.1$ and $\lambda=-1$ shown in figure~\ref{fig:totalcounter01}.
However, firstly, there is the downward spike in this case. Secondly,
it appears in the retrograde neutrino scattering depicted in figures~\ref{fig:4b}
and~\ref{fig:4d}. This feature results from the enhancement
of the transition probability $P_{\mathrm{LR}}^{(\mathrm{tot})}$
for ultrarelativistic neutrinos with some impact parameter. Performing
numerical simulations, we have obtained that the border between the
qualitatively different cases, shown in figures~\ref{fig:totalcounter05}
and~\ref{fig:totalcounter01}, is at $z\approx(0.12-0.13)$.

The explanation of the features in figures~\ref{fig:3a}
and~\ref{fig:4b} is not obvious. Unlike neutrino
oscillations in constant external fields, where a resonance in oscillations
can be revealed just from the expression of an effective Hamiltonian,
here the external fields in eq.~(\ref{eq:Omegax}) are coordinate
dependent. Moreover, this effect is integral, i.e. we have to make
a neutrino to pass on the whole trajectory and, then, obtain the state
of the polarization of the particle. Therefore, in our case, one should
rely mainly on the numerical simulations rather than on the naive
analysis of the effective Hamiltonian in eq.~(\ref{eq:Schreq}).

Nevertheless, the appearance of spikes in figures~\ref{fig:3a} and~\ref{fig:4b} can be accounted for qualitatively by the neutrino interaction with moving matter. Plasma of an accretion disk moves with relativistic velocities on circular orbits; see eq.~\eqref{eq:Uftr}, where $U_{f}^{\phi} \neq 0$. In the neutrino gravitational scattering, there are situations when the neutrino velocity has a nonzero angle with respect to the plasma velocity. Hence, the term $\propto G_\mathrm{F}\mathbf{g}$ in $\bm{\bm{\Omega}}$ in eq.~\eqref{eq:vectG} has a component perpendicular to the neutrino velocity. It is this term, which causes the neutrino spin-flip. Analogous effect in neutrino (flavor) oscillations was considered in ref.~\cite{GriLobStu02}.

\section{Conclusion}

We have studied spin effects in the neutrino scattering off a rotating
SMBH. This SMBH is supposed to be surrounded by a realistic magnetized
accretion disk. We assume that neutrinos are ultrarelativistic particles.
These neutrinos are supposed to interact with dense matter of the
accretion disk by the electroweak forces. We use the forward scattering
approximation in the neutrino interaction with matter. It means that
the matter contribution is in the invariant four vector effective
potential $G^{\mu}$. The neutrino interaction with the magnetic field
in a disk is owing to the presence of a nonzero diagonal neutrino
magnetic moment, i.e. we assume that a neutrino is a Dirac particle.
Of course, we exactly account for the effects of curved spacetime
on the propagation and spin oscillations of neutrinos.

Considering the equatorial motion of neutrinos in section~\ref{sec:SPINEVOL},
we have derived the effective Schr\"odinger equation for the evolution
of the neutrino polarization in the particle scattering. Initially
all ultrarelativistic neutrinos are taken to be left polarized. A detector
is also sensitive to left polarized neutrinos. Thus, the measured
flux is the product of the survival probability of spin oscillations
and the outgoing flux of scalar particles. The latter flux is computed
by the standard technique (see, e.g., ref.~\cite{DolDorLas06}).

In section~\ref{sec:APPL}, we have chosen the parameters of the accretion
disk, such as the radial distributions of background matter and the
magnetic field, and the value of the neutrino magnetic moment. These
parameters are realistic and do not violate current constraints. We
have also found in section~\ref{sec:APPL} that solely gravitational
interaction does not contribute to the spin-flip in the scattering
of ultrarelativistic neutrinos in the Kerr metric. This our result
is in agreement with the finding of ref.~\cite{Lam05}. However,
we generalized the statement of ref.~\cite{Lam05} by considering
arbitrary neutrino trajectories.

The outgoing fluxes of neutrinos based on the numerical solution of
eq.~(\ref{eq:Schreq}) are presented in section~\ref{sec:RES}. If
we neglect the contribution of the accretion disk, eq.~(\ref{eq:Schreq})
can be solved in quadratures. The corresponding fluxes for the direct
and retrograde scatterings are shown in figure~\ref{fig:magnetic}.
Spin effects are negligible for the forward neutrino scattering: $F_{\nu}(\chi=0)=F_{0}$.
The maximal deviation of the flux is for the backward neutrino scattering
at $\chi=\pi$. It can reach about 6\%.

Then, we have taken into account the neutrino interaction with matter.
We have considered accretion disks which corotate and counter-rotate
SMBH. The case of a corotating disk differs from the magnetic case
only quantitatively; cf. figures~\ref{fig:magnetic} and~\ref{fig:totalcorot}.
However, if a disk counter-rotates ($\lambda=-1$), there are spikes
in the total outgoing fluxes. These spikes appear either in the direct
scattering, see figure~\ref{fig:3a}, or in the retrograde
scattering, see figure~\ref{fig:4b}, depending on the
angular momentum of BH $J=2M^{2}z$. One can treat these spikes, especially
that shown in figure~\ref{fig:4b}, as the appearance
of a quasi-resonance in spin oscillations.

The advance of the present work in comparison with refs.~\cite{Dvo20a,Dvo20b}
is the following. Firstly, we have accounted for the neutrino electromagnetic
interaction with the magnetized accretion disk. Secondly, we have
improved the description of the accretion disk. In refs.~\cite{Dvo20a,Dvo20b},
background matter was taken to be distributed isotropically around
BH. Now, we suppose that particles of plasma rotate around BH on circular
orbits, with the four velocity given in eq.~\eqref{eq:Uftr}. Hence
we consider a realistic accretion disk. We have also taken the indexes
in the matter distribution $\beta$ and the magnetic field radial
dependence $\kappa$, which result from the equipartition of the energy
between the magnetic field and the accreting plasma. Finally, we have
corrected the statement of ref.~\cite{Dvo20b} on the gravity contribution
to the spin-flip of ultrarelativistic neutrinos scattering in the
Kerr metric.

There is an assumption, which significantly simplifies the calculations
in the present work: we have studied neutrinos moving in the equatorial
plane only. Neutrinos in the vicinity of BH can be emitted inside
its accretion disk in nuclear reactions between plasma particles.
Then, these neutrinos are lensed, with their spin precessing in external
fields in curved spacetime. In the present work, we have studied a
relativistic accretion disk. It means that neutrinos are emitted mainly
along the velocities of background fermions in an accretion disk, i.e. the transverse
momentum of neutrinos is much less than the longitudinal one. Hence,
the approximation of the equatorial neutrino motion is valid. Note
that the arbitrary trajectories of massless particles (photons) emitted
by an accretion disk of SMBH were studied in ref.~\cite{DokNaz20}.

In summary, we have studied the neutrino scattering off a rotating
SMBH surrounded by a magnetized accretion disk with the realistic
matter density distribution. Considering the conservative value of
the neutrino magnetic moment $\mu=10^{-14}\,\mu_{\mathrm{B}}$ and
the moderate magnetic field strength in the vicinity of SMBH $B_{0}\sim10^{2}\,\text{G}$~\cite{Dal19},
we have obtained that the outgoing neutrino flux can be reduced by
$(3-6)\%$. Although the predicted phenomenon is beyond the sensitivity of the current neutrino telescopes, one expects that it can be observed in future (see, e.g., ref.~\cite{Abe18}).


\acknowledgments

I am thankful to P.M.~Akhmet'ev, A.N.~Baushev, and K.A.~Postnov
for useful discussions. The work is supported by the government assignment
of IZMIRAN.

\appendix

\section{An electromagnetic field near BH\label{sec:EMBH}}

In this appendix, we briefly remind how to construct the electromagnetic
field in the vicinity of a Kerr BH which asymptotically reaches the
constant and uniform magnetic field $B$. Basically, we follow ref.~\cite{Wal74}.

The Kerr metric in eq.~(\ref{eq:Kerrmetr}) has two Killing vectors:
$\eta^{\mu}=(1,0,0,0)$ and $\psi^{\mu}=(0,0,0,1)$. They satisfy
the relation, $\psi_{\mu;\nu}+\psi_{\nu;\mu}=0$ and analogously for
$\eta_{\mu}$. The one-form $\psi=\psi_{\mu}\mathrm{d}x^{\mu}$ corresponding
to the Killing vector $\psi^{\mu}$ generates the electromagnetic
field $F_{\psi} \propto \mathrm{d}\psi=\tfrac{1}{2}(\nabla_{\mu}\psi_{\nu}-\nabla_{\nu}\psi_{\mu})\mathrm{d}x^{\mu}\wedge\mathrm{d}x^{\nu}$.
One can check that this electromagnetic field obeys the Maxwell equation,
$(F_{\psi})_{\quad;\nu}^{\mu\nu}=0$. Analogous contribution results
from $\eta=\eta_{\mu}\mathrm{d}x^{\mu}$.

Thus, the total electromagnetic field has the form,
\begin{equation}
  F=a_{\psi}\mathrm{d}\psi+a_{\eta}\mathrm{d}\eta,\label{eq:F2form}
\end{equation}
where the coefficients $a_{\psi,\eta}$ are uniquely fixed by the
two conditions. Firstly, the total electric charge of BH equals zero, i.e.
$\smallint*F=0$, where the integration is over the 2D surface with
$t=\text{const}$ and $r\to\infty$. Accounting for the facts that~\cite{CohWal72}
$\smallint*\mathrm{d}\psi=-16\pi Ma$ and $\smallint*\mathrm{d}\eta=8\pi M$,
we get that $a_{\eta}=2aa_{\psi}$.

Secondly, asymptotically, i.e. at $r\to\infty$, the components of
the electromagnetic field $F$ in eq.~(\ref{eq:F2form}) should be
$F_{r\phi}=-F_{\phi r}=-Br\sin^{2}\theta$ and $F_{\theta\phi}=-F_{\phi\theta}=-Br^{2}\sin\theta\cos\theta$.
It corresponds to the uniform magnetic field $B$ along the rotation
axis of BH. It gives one $a_{\psi}=B/2$ and $a_{\eta}=aB$.

We can obtain the vector potential of the electromagnetic field $F=\mathrm{d}A=\tfrac{1}{2}(\nabla_{\mu}A_{\nu}-\nabla_{\nu}A_{\mu})\mathrm{d}x^{\mu}\wedge\mathrm{d}x^{\nu}$
in the form, $A_{\mu}=\tfrac{B}{2}(\psi_{\mu}+2a\eta_{\mu})=\tfrac{B}{2}(g_{\mu\phi}+2ag_{\mu t})$.
Using eq.~(\ref{eq:Kerrmetr}), one finds that it has two nonzero
components,
\begin{equation}\label{eq:Atphi}
  A_{t}=Ba
  \left[
    1-\frac{rr_{g}}{2\Sigma}(1+\cos^{2}\theta)
  \right],
  \quad
  A_{\phi}=-\frac{B}{2}
  \left[
    r^{2}+a^{2}-\frac{a^{2}rr_{g}}{\Sigma}(1+\cos^{2}\theta)
  \right]\sin^{2}\theta,
\end{equation}
which are used in section~\ref{sec:SPINEVOL} to build the electromagnetic
field in the world coordinates, $F_{\mu\nu}=\nabla_{\mu}A_{\nu}-\nabla_{\nu}A_{\mu}$.

\end{document}